\documentclass{aa}

\usepackage{graphics}

\begin{document}

\thesaurus{08(08.02.1;08.09.2 GRO J1655-40; 02.01.2;02.09.1;13.25.3}


\def\aaa#1{{A\&A,} {#1}}
\def\acta#1{{Acta\ Astr.,} {#1}}
\def\aas#1{{A\&AS,} {#1}}
\def\annrev#1{{ARA\&A,} {#1}}
\def\aj#1{{AJ,} {#1}}
\def\apj#1{{ApJ,} {#1}}
\def\apjs#1{{ApJS,} {#1}}
\def\mnras#1{{MNRAS,} {#1}}
\def\nat#1{{Nature,} {#1}}
\def\pasj#1{{PASJ,} {#1}}
\def\pasp#1{{PASP,} {#1}}
\def\phylet#1{{Phys.\ Lett.\ A.,} {#1}}
\def\ptp#1{{Prog.\ Theor.\ Phys.,\/} {#1}}
\def\sci#1{{Science,} {#1}}
\def\sp#1{{Solar\ Phys.,} {#1}}
\def\ssr#1{{Space\ Sci.\ Rev.,\/} {#1}}

\def\J16{{GRO~J1655--40}}
\def\Msun{\rm M_{\odot}}
\newcommand{\gta}{\mathrel{\hbox{\rlap{\lower.55ex \hbox {$\sim$}}
                   \kern-.3em \raise.4ex \hbox{$>$}}}}
\newcommand{\lta}{\mathrel{\hbox{\rlap{\lower.55ex \hbox {$\sim$}}
                   \kern-.3em \raise.4ex \hbox{$<$}}}}


\title{The 1996 outburst of GRO J1655-40: disc irradiation and \\
enhanced mass transfer}

\author{A.A. Esin\inst{1,2}\thanks{Chandra Fellow}\and
J.-P. Lasota\inst{1,3}\thanks{also: DARC, Observatoire de Paris, France}
\and R. I. Hynes\inst{4,5}}
\institute{Institute for Theoretical Physics, University of California, Santa Barbara CA 93106-4030, USA \and
Theoretical Astrophysics, 133-30 Caltech, 
Pasadena CA91125, USA; E-mail: aidle@tapir.caltech.edu \and 
Institut d'Astrophysique de Paris; CNRS, 98bis Bd. Arago, 75014 Paris, France; E-mail: lasota@iap.fr \and
Department of Physics and Astronomy, The Open University, Walton Hall, 
Milton Keynes, MK7 6AA, UK; \and
Department of Physics and Astronomy, University of Southampton, 
Hampshire SO17 1BJ, UK; E-Mail: rih@astro.soton.ac.uk}

\offprints{Esin}
\mail{Lasota}

\date{Received/Accepted}

\titlerunning{The outburst of \J16}
\authorrunning{A.A. Esin et al.}

\maketitle

\begin{abstract}
We show that the 1996 outburst of the X-ray binary transient system
\J16 can be explained by the standard dwarf-nova type disc
instability, followed by an episode of enhanced mass transfer from the
secondary if the mass transfer rate in \J16 is within a factor $\lta
10$ of the stability limit. We argue that irradiation of the
secondary during the onset of the outburst driven by the thermal
instability in the outer disc can increase the mass transfer
rate above the minimum value required for stable accretion. This will
then produce the period of near-constant X-ray emission seen in
this system. This scenario can also explain the observed
anti-correlation between the optical and X-ray fluxes.  It is
generally accepted that optical emission in low-mass X-ray binaries is
produced by irradiation of the outer disc by X-rays.  There is also
strong circumstantial evidence that in order for the outer disc to see
the irradiating flux, it must be warped. Depending on the warp
propagation mechanism, either a burst of mass from the secondary or
viscous decay are likely to decrease the degree of warping, thereby
causing the decrease in the observed optical flux while the X-ray flux
remains constant or even increases, exactly as observed in
\J16. Finally, the decrease of the disc warping and, therefore,
irradiation will cause the disc to become unstable once again,
terminating the outburst.
\end{abstract}

\section{Introduction}
\label{intro}

The X--ray source \J16 has several unusual characteristics.  Though,
like all soft X-ray transients, it is a low-mass binary system, its
1.7--$3.3 \Msun$ secondary orbiting a 5.5--$7.9 \Msun$ black hole
(Shahbaz et al. 1999) is considerably more massive than is typical in
such systems (see e.g. a review by Tanaka \& Shibazaki 1996). Its mass
and spectral type (F3-F4) imply that the donor star is  near (or
just beyond) the end of its main-sequence life-time.  In fact Kolb et
al. (1997) and Kolb (1998) assert that it appears to be crossing the
Hertzsprung gap, i.e. expanding towards the giant branch, which would
imply a very large mass transfer rate, $\dot M_{\rm T} \sim 10^{19}$ g
s$^{-1}$ (King \& Kolb 1999). Even if the secondary star in \J16 is
still on the main sequence, as argued by Reg\"os, Tout \&
Wickramasinghe (1998), the inferred mass transfer rate, $\dot M_{\rm
T} \gta 10^{17}$ g s$^{-1}$, is orders of magnitude greater than $\dot
M_{\rm T}$ estimates in other X-ray transients (see e.g. Chen, Shrader
\& Livio 1997).  Moreover, the values quoted above are close to the
critical rate, above which the irradiated accretion disc in \J16 would
be stable with respect to the dwarf-nova type instability (van
Paradijs 1996; see Dubus et al. 1999b for a most recent discussion of
the critical rate for this instability), that is generally thought to
be responsible for the transient behaviour of low-mass X-ray
binary systems.

The outbursts themselves are also rather atypical. Like other Black
Hole (Low Mass) X-ray Transient systems (BHXTs), \J16 has been
quiescent for more than 30 years before entering the active phase in
1994. However, contrary to the behaviour of `conventional' BHXTs, the
first outburst was followed in 1995 by two others which displayed hard
X-ray spectra.  Finally, after the last burst of emission in
July/August 1995 the system settled into X--ray quiescence, with
luminosity of $L_X\approx 2 \times 10^{32}$ erg s$^{-1}$.

This quiescent state ended around April 25, 1996  when a new, soft
X--ray outburst began, which is the main subject of this paper.
Orosz et al.\ (1997) obtained BVRI photometry close to the onset
of the outburst and found the rise in X-ray flux delayed by 6 days
with respect to the optical increase.
Consequently, Hameury et al. (1997) showed
that the rise to outburst, and in particular the X-ray
`delay', are very well described by the dwarf-nova type disc
instability model (DIM) if the accretion disc is truncated at $\sim 5
\times 10^3$ $R_S$ ($R_S=2GM/c^2$ is the Schwarzschild radius).
However, the success of this model is put into perspective by the
subsequent behaviour of the system during the outburst, when, after the
first local maximum, the soft X-ray light curve rose to a higher
luminosity `flaring' plateau. The present theoretical understanding
tells us that this type of light-curve cannot be produced based on the
standard DIM (Hameury et al. 1998).

The behaviour of the optical luminosity during the 1996 outburst of
\J16 also seems to defy the generally accepted idea (based on a solid
observational background) that in Low Mass X-ray Binaries (LMXBs) the
optical emission from the accretion disc is due to reprocessing of
X-rays in the outer disc. The fact that in \J16 the optical flux 
decreases while the X-ray flux increases and then fluctuates around an
approximately constant value is difficult to reconcile with the X-ray
reprocessing model (Hynes et al. 1998).
Finally, the list of unusual properties of \J16 should be completed by
the presence of a superluminal `jet' seen during the 1994
outburst (Hjellming \& Rupen 1995).

In this paper we argue that the puzzling behaviour of this system can
be readily reconciled  with the DIM, if we take into account the
fact that the mass transfer rate in \J16 is likely to be rather close to
the stability limit.  In this case, irradiation of the secondary
during an outburst may be enough to increase the mass transfer rate by
a factor of a few and thereby push the system into the stable regime.
This scenario and its expected effect on the observed X-ray light
curve are discussed in detail in \S\ref{xray}.  Furthermore, in
\S\ref{opt} we speculate that the reason the optical flux does not
rise with increasing X-ray luminosity is a reduction in the
amplitude of the outer disc warping. Using a flared planar disc
`approximation' to describe a warped disc, we show that as a result,
the intercepted X-ray flux and therefore the observed optical emission
is reduced.  We further suggest in \S\ref{end} that the flattening of
the outer disc and a consequent decrease in X-ray irradiation may also
be responsible for the ultimate end of the stable accretion phase, by
raising the stability criterion.  We conclude with a final discussion
and a summary in \S\ref{disc}.

\section{The X--ray Outburst}
\label{xray}

Following Shahbaz et al. (1999), in our estimates below we assume that \J16
contains a 7$\,\Msun$ black hole and that the mass ratio between the
primary mass $M_1$ and the secondary mass $M_2$ is $q\equiv M_2/M_1 =
0.33$.  We also adopt a distance to the system of 3.2$\,{\rm kpc}$
(Hjellming \& Rupen 1995).

\begin{figure}
\resizebox{\hsize}{!}{\includegraphics{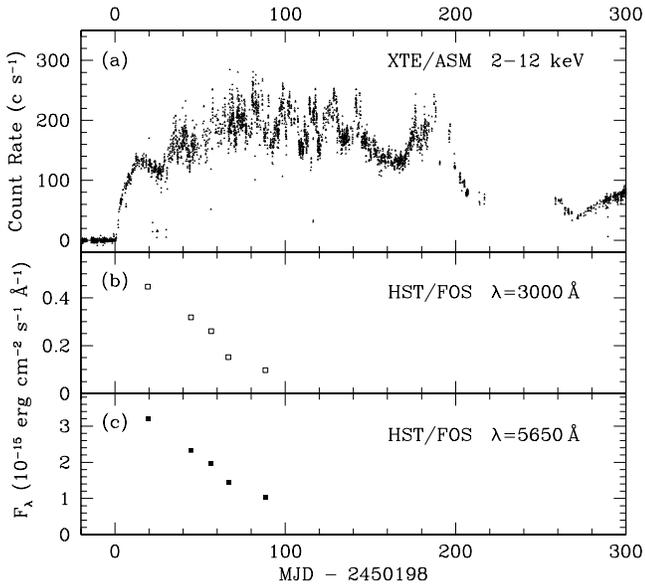}}
\caption{X-ray (a), UV (b) and optical (c) light-curves (Sobczak et
al. 1999; Hynes et al. 1998) of the 1996 outburst of \J16.  A
comparison of the  three panels clearly shows anti-correlation between
the observed optical (and UV) fluxes and soft X-ray count rates.  
The data on panels (b) and (c) has not been corrected for
interstellar reddening.  Note that the flux scales
in panels (b) and (c) supersede those on Fig.\ 3 of Hynes et al.\
(1998) which were in error.}
\label{ltcrv}
\end{figure}

Figure \ref{ltcrv}(a) shows the progress of the 1996 X-ray outburst of
\J16.  The ASM (2-12 keV) `soft' X-ray flux began to rise 6 days after
the start of the optical outburst and 15 days later attained a maximum
corresponding to $\sim 0.12 L_{\rm Edd}$ (hereafter we use spectral
fits by Sobczak et al. 1999 to convert ASM fluxes to bolometric luminosity
values), where $L_{\rm Edd}= 1.25\times 10^{38} {M/\Msun} \ {\rm erg \
s^{-1}}$ is the Eddington luminosity.  After $\sim 12$ days of a
roughly exponential decline which followed the first maximum, the soft
X-ray flux fell by about $30\%$.
It then began to rise again, and after about two months from the onset
of the outburst, reached a strongly flaring plateau with {\sl total}
luminosity varying around $\sim 0.17 L_{\rm Edd}$.  This flaring
state, which Sobczak et al. (1999) identify with the very high
spectral state, continued roughly until day 200 of the X-ray outburst.
Over the next $\sim 70$ days the luminosity fell to less than a third
of its peak value.  Note that this local minimum was followed by a
$\sim 150$ day long reflare, which we  do not try to address in
this paper (but see \S\ref{end}).  Our goal here is to apply the
(modified) DIM to the main 1996 outburst of \J16 since, as discussed
above, this model was successfully used to describe the beginning of
this event.  Various type of `reflares' (observed also in dwarf novae)
do not have a clear explanation in the context of the DIM (see e.g.
Hameury, Lasota \& Warner 2000).  One can speculate that the
long rise-time of the \J16's reflare could be due to an inside--out
outburst from a non--truncated disc, while its flat-topped
lightcurve suggests a quasi--steady accretion phase.  However, in view
of incomplete optical coverage and uncertainties in models
discussed below, we feel that an attempt at serious modeling of the
reflare would be pure guesswork.

It is very suggestive that the e-folding rise time of $\lta 1$ day and
the e-folding decay time of $\sim 35$ days, observed in the first 27
days of the X-ray outburst of \J16, are characteristic of the
so-called FRED-type light curves seen in many X-ray transients (see
e.g. Chen, Shrader \& Livio 1997).  The rise and decay (see
Figs. \ref{ltcrv}[b] and \ref{ltcrv}[c]) of the optical and UV
luminosity also follow the pattern of a FRED outburst with an
e-folding decline time close to $\sim 73$ days, considerably longer
than that for X-rays.  This behaviour naturally follows from the DIM
in which the outer regions of the accretion disc are irradiated by the
X-rays from the inner region (King \& Ritter 1998; Dubus, Lasota \&
Hameury 1999a). One could conclude, therefore, that \J16 was producing
a `normal' X-ray transient outburst when, on day 27 a new event
occurred which stopped the decline of the X-ray flux.  In what
follows, we will argue that the change in the outburst pattern was due
to an increase in the mass transfer rate, which moved the system
parameters into the range corresponding to {\sl stable} accretion for
an X-ray irradiated disc.

The minimum mass transfer rate necessary for a stable, steady-state accretion
in an X-ray irradiated disc may be written as (Dubus et al. 1999b)
\begin{eqnarray}
\dot M^{\rm irr}_{\rm crit} &\approx& 2.8 \times 10^{18} \left({R_{\rm out} 
\over 5.3 \times 10^{11} {\rm cm}}\right)^{2.1} \times \nonumber \\ 
&&\left(\frac{\cal C}{5 \times 10^{-4}}\right)^{-0.5} ~\rm g~s^{-1}.
\label{mcrirr}
\end{eqnarray}
Here the phenomenological `parameter' (in reality a function of radius
and time, see \S\ref{opt}) $\cal C$ provides a simple description of the
disc irradiation properties through:
\begin{equation}
T^4_{\rm irr} = {\cal C} {\dot M c^2 \over 4 \pi \sigma R^2},
\label{tirr2}
\end{equation}
where $T_{\rm irr}$ is the irradiation temperature. ${\cal C} = 5
\times 10^{-4}$ is the value that represents best the average
irradiation flux intercepted by the the outer disc regions in LMXBs
(see Dubus et al. 1999b\footnote{Dubus et al. write that this
value is used `by comparison with a formula extensively used
in the literature' but in fact the numerical value of ${\cal C}$ 
is deduced from
applications of this formula to observations.}).  
We assumed that in outburst the disc outer radius
$R_{\rm out}$ expands to  $90\%$ of the primary Roche lobe
radius (Smak 1999b), so that
\begin{equation}
R_{\rm out}= 5.3 \times 10^{11} \left({M_1\over 7\Msun}\right)^{1/3} {\rm cm}
\label{rdisc}
\end{equation}
Let us note, however, that if the outward propagating heat front 
(Hameury et al. 1998) did not reach the outermost disc regions, the  
disc expansion would be negligible, and its outer edge would extend only to 
$\sim 70\%$ of the Roche lobe radius.

The critical accretion rate given by Eq. (\ref{mcrirr}) is very 
uncertain, as discussed in Dubus et al. (1999a), and should be considered 
an order of magnitude estimate only.
For the accretion in an irradiated disc to be
stable the mass-transfer rate from the secondary has, therefore, to be
larger than $\sim 10^{18}~{\rm g~s^{-1}}$.  For \J16 this corresponds
to $\dot M_{\rm T} \sim 0.1 \dot M_{\rm Edd}$ ($\dot M_{\rm
Edd}=L_{\rm Edd}/0.1 c^2$). 

Since $L \sim 0.2 L_{\rm Edd}$ during the plateau phase of the
outburst, the observations are entirely consistent with a {\em stable}
disc accreting at $\dot M = \dot M_{\rm T} \sim 0.2 \dot M_{\rm
Edd}$. For our arguments to work, however, the mass transfer rate from
the secondary must be smaller than critical value of $0.1 \dot M_{\rm
Edd}$ quoted above, since otherwise \J16 would not be a transient, at
least according to the DIM. So what is known about $\dot M_{\rm T}$ in
this system?

Estimates of the mass transfer rate in \J16 quoted in the literature
differ by three orders of magnitude.  At the high end, Kolb et
al. (1997) and King \& Kolb (1999) argue that the secondary star in
\J16 is crossing the Hertzsprung gap between the main sequence and the
red giant branch, and is therefore undergoing a rapid envelope
expansion (on a time scale $\sim 10^7$ years).  They conclude that
this implies a mass transfer rate of order $M_2/(10^7 {\rm y}) \gta
10^{19} {\rm g~s^{-1}}$.  Obviously, taken at face value, this
result suggests that \J16 cannot be a transient system.  Reg\"os et
al. (1998) show, however, that the type F3-F4 donor can still lie on
the main sequence if convective overshooting at the core--envelope
boundary is taken into account in stellar modeling. The presence of
such (or equivalent) mixing is expected for main--sequence stars in
the relevant mass range (see Reg\"os et al. 1998 and references
therein).  Based on this argument, Reg\"os et al. (1998) derive a
considerably lower value for the mass transfer rate in \J16,
estimating that $\dot M_{\rm T} \sim 10^{17} -10^{18}\,{\rm g~s^{-1}}
\approx 0.1 - 1.0 \times \dot M^{\rm irr}_{\rm crit}$.  If this
estimate is correct, the quiescent accretion disc in \J16 is still
unstable, though with the mass transfer rate rather close to the
critical one.  In this case, an increase in $\dot M_{\rm T}$ by a
factor of a few during the outburst would stabilize the accretion, and
produce the `plateau' in the light curve, just as observed during the
1996 outburst.  Interestingly, Hameury et al. (1997) showed that the
spectrum of \J16 in quiescence as well as its rise to outburst in 1996
can be very well modeled using the standard DIM operating in a
truncated disc, assuming that the mass transfer rate is $2\times
10^{17} {\rm g~s^{-1}}$, within the range considered plausible
by Reg\"os et al. (1998).

These relatively high values of $\dot M_{\rm T}$ are somewhat
difficult to reconcile with the estimate of van Paradijs (1996), who
used the standard method of deducing the mass transfer rates in X-ray
transients by dividing the mass accreted during the outburst by the
recurrence time. By estimating the total emission observed during the
1994--1995 period of activity and assuming a recurrence time of $>30$
years van Paradijs (1996) obtained $\dot M_{\rm T} \lta 8 \times
10^{15} {\rm g~s^{-1}}$ (see also Menou et al. 1999). However, the
validity of this method for the recent outburst cycle of \J16 is
questionable, since the observed outbursts clearly do not occur with
even rough regularity. Between August 1995 and April 1996 \J16 was in
true quiescence so the recurrence time for the 1996 outburst was only
9 months. The relevant mass transfer rate may then be estimated by
dividing the mass accreted during the 1996 outburst (which was roughly
$4 \times 10^{24} {\rm g}$, assuming the standard $10\%$ radiative
efficiency) by this recurrence time. The result, $\dot M_{\rm T}
\approx 2 \times 10^{17} {\rm g~s^{-1}}$,  is in reasonable
agreement with convective overshooting models of Reg\"os et
al. (1998). This, of course, does not resolve the problem of the
pre--1994 phase, namely that no outbursts were observed during the
preceding 30 years.

Overall we conclude that there is significant evidence pointing to the
mass transfer rate in \J16 being rather {\em close} to the
stability limit of an irradiated disc, at least during the period
directly preceding the 1996 outburst.  Furthermore, we can
confidently state that the transient nature of this system as well as
the shape of the X-ray light curve seen in 1996 can be attributed to
the dwarf--nova type disc instability only if $\dot M_{\rm T}$ in
quiescence was below, but not too far from the critical mass transfer
rate.  The former condition is necessary to have outbursts at all, and
the latter is desirable so that a reasonable increase in $\dot M_{\rm T}$
during the first part of the outburst is sufficient to stabilize the disc.
In addition, the same value of $\dot M_{\rm T}$ is required to explain
all the rise--to--outburst properties.

An obvious reason for an increase in $\dot M_{\rm T}$ is X-ray
irradiation of the secondary star. According to Phillips, Shahbaz \&
Podsiadlowski (1999), X-ray heating strongly affects the vertical
structure of the irradiated outer layers of the secondary. Assuming an
X-ray luminosity of $1.4 \times 10^{37}$ erg s$^{-1}$, they obtain for
\J16 a (maximum) ratio of the irradiating to the intrinsic stellar
flux $\sim 6.6$. They were considering BATSE (20 -200 keV) X-ray
fluxes observed during the March 1995 event, whereas during a typical
outburst most of the energy is emitted in softer (2-20 keV) X-rays
(Sobczak et al. 1999). At the first maximum in the 1996 outburst, the
total X-ray luminosity was approximately 8 times larger than the value
used by Phillips et al. (1999), so the flux ratio would instead
be $\sim 50$ and the expected effect considerably larger. 
Thus, an increase of the mass-transfer rate
due to irradiation should be expected. In addition, in a recent
seminal article, Smak (1999a) showed that mass--transfer enhancements
play a fundamental role in dwarf--nova outbursts (see \S\ref{disc})
and therefore they should also be expected in X--ray transients.

As pointed out by Phillips et al. (1999) X-ray irradiation does not
directly affect the vicinity of the $L_1$ point but heats up matter
located higher in altitude. This will induce a delay, because the
heated matter has to move to the Roche nozzle before falling onto the
accreting object.  Such a delay was observed in dwarf novae; when the
heating of the secondary by UV and optical emission is clearly
observed, brightening of the hot spot due to the increased mass
transfer appears a few days after the outburst maximum (Smak 1995).
If the increased mass-transfer rate is due to irradiation, this delay
must be shorter than the disc viscous time, in order to keep $\dot
M_{\rm T}$ approximately constant. In the opposite case, the
light-curve would instead have an exponential shape (Hameury, King \&
Lasota 1988; Augusteijn, Kuulkers \& Shaham 1993; Hameury, Lasota \&
Hur\`e 1997). 

The disc viscous time is
\begin{eqnarray}
t_{\rm vis} &\approx& {R^2 \over \nu}= \
320  \left({\alpha\over 0.1}\right)^{-4/5}
\left({\dot M \over 10^{18} {\rm g \ s^{-1}}}\right)^{-3/10}  
\times \nonumber \\
&&\left({M \over 7 \Msun}\right)^{1/4}
 \left({R \over 10^{11}{\rm cm}}\right)^{5/4} {\rm d},
\label{tvis}
\end{eqnarray}
using the expression for midplane temperature given for
the `hot' Shakura \& Sunyaev (1973) solution. At \J16's outer disc
radius this time is longer than the total duration of the 1996
activity ($\sim 500$ days), at least for values of $\alpha \approx
0.1$ - 0.2 assumed in the DIM.

In our picture, irradiation of the secondary at the time of the first
maximum in the X-ray light curve, would increase the mass
transfer rate from the companion.  To be consistent with observations,
this increase should affect the accretion rate onto the black hole
after about two weeks, a considerably shorter time period than $t_{\rm
vis}$ above.  In this case, however, the relevant time scale is not
that given by Eq. (\ref{tvis}), but instead the time it takes for a surface
density front, created by a sudden increase in mass transfer, to
diffuse towards the inner disc regions.  In the presence of a warp, matter 
transfered from the companion is fed to the disc not at the outer edge but 
at radius $\sim R_{\rm circ}/2$ (see e.g. Wijers \& Pringle 1999), where 
\begin{equation}
R_{\rm circ} \approx 2.13 \times 10^{11} \left({M_1\over 7\Msun}\right)^{1/3} 
{\rm cm}
\label{rcirc}
\end{equation}
is the so called circularization radius corresponding to the specific
angular momentum of the mass transferred through the $L_1$ Lagrangian 
point.
The time in which this surface density excess will reach the black
hole is then given by (Hameury et al. 1997)
\begin{equation}
t_{\rm vis}^c \lta t_{\rm vis}(R_{\rm circ}) {\delta R \over R_{\rm circ}}
\label{tvisd}
\end{equation}
where $\delta R$ is the width of the surface density contrast. A
density excess with $\delta R/R_{\rm circ} \sim 0.1$ 
(see e.g. Frank et al. 1987) would diffuse to
the central black hole in $\sim 20$ days 
in agreement with our scenario.   As we pointed out above, this
number should be considerably longer than the expected delay between
irradiation of the secondary and the increase in mass transfer, as
required to produce a flat light-curve. Although it is very difficult
to estimate how long it takes for the heated matter to reach the $L_1$
point (see Hameury et al. 1993, for discussion of a related problem),
a time of $\sim$ few days is reasonable in the sense that it would
imply subsonic speeds.

Our explanation of the `flat top' X-ray light-curve of \J16 is very
similar to the one proposed by King \& Cannizzo (1998) for
light-curves of Z Cam-type dwarf-nova systems (see their
Fig. 2). These authors do not invoke irradiation to explain the
increased mass-transfer rate, however, but attribute mass
transfer variations to a starspot.

Interestingly, the King \& Cannizzo (1998) model is more successful in our
case than it is in explaining Z Cam `standstills'. 
In Z Cam the luminosity is observed to stick at a constant level halfway
down from maximum, whereas in light curves produced by King \& Cannizzo the
the `standstill' is
at a luminosity {\sl higher} than the outburst maximum, as in
\J16. King \& Cannizzo expect this to be the result of the high
mass-transfer enhancement factor (6) used in their calculation. In
reality, however, the mass-transfer rate corresponding to stable
disc accretion is, by construction, {\sl always} higher than
the maximum outburst mass accretion rate. The reason
(see e.g. Hameury et al. 1998) is that at outburst
maximum, before the cooling wave begins to propagate, the accretion
rate in the disc is almost constant, i.e. at the outer disc edge (or at the
hot disc outer edge) it is close to the accretion rate corresponding
to the critical surface density ($\Sigma_{\rm min}$ marking the end of
the `hot' branch of the S-curve representing disc equilibria). A
stable mass-transfer rate must be larger than this value.

\section{The Optical Light Curve}
\label{opt}

\begin{figure}
\resizebox{\hsize}{!}{\includegraphics{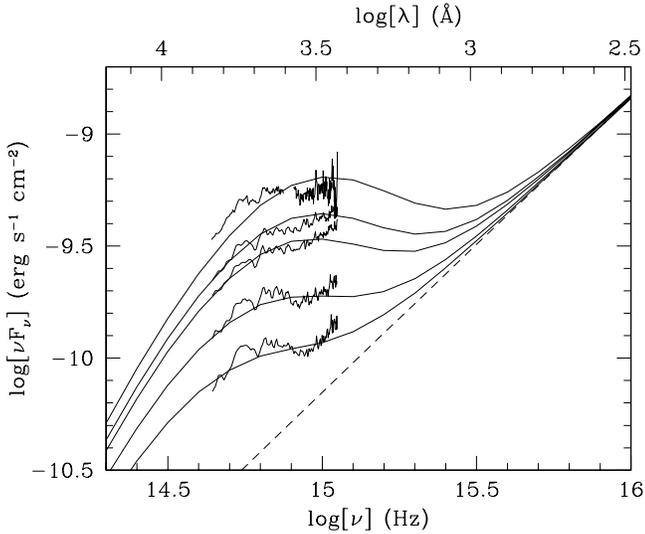}}
\vskip -1.5cm
\caption{The observed optical/UV spectra of \J16 are plotted together
with model spectra from a flared disc irradiated by X-rays emitted in
the inner parts of the disc.  The data were corrected for interstellar
reddening, assuming $E(B-V) =1.2$; we also subtracted the contribution
from the secondary.  The model spectra were calculated for $\dot M =
0.2 M_{\rm Edd}$ and disc size given by Eq. (\ref{rdisc}).  Solid
curves correspond to different values of the irradiation parameter at
the outer disc edge, ${\cal C}_{\rm out}$; top to bottom ${\cal
C}_{\rm out} = 6\times 10^{-4}, 4\times 10^{-4}, 3\times 10^{-4},
2\times 10^{-4}, 0.8\times 10^{-4}.$ Dashed curve shows the spectrum
of a planar (non-irradiated) disc.}
\label{spectra}
\end{figure}

There is now a general consensus that optical emission in persistent
and transient (in outburst) low-mass X-ray binaries is due to
reprocessing of X-rays by the outer disc (see e.g. van Paradijs \&
McClintock 1995).  This is also the case for \J16, since a simple
estimate of the expected optical flux from a non-irradiated disc which
reproduces the observed X-ray emission falls short of the observed
optical flux by roughly an order of magnitude (see
Fig. \ref{spectra}).  Moreover, as we pointed out in \S\ref{xray}, the
optical light-curve is entirely consistent with the decline from
maximum during a typical FRED-type outburst in an irradiated disc.
Longer UV and optical decline time scales ($\sim 73$ days, as
compared to $\sim 35$ days for the X-ray flux) are simply due to the
fact that as the irradiating flux declines, the outer disc edge
becomes cooler and the peak of the emission moves into the optical
band, thus compensating for the decrease in the total emission from
the outer disc.

However, this simple picture clearly cannot explain the behaviour of
the optical flux at times later than $\sim 30$ days after the onset of
X-ray outburst.  At this time, X-ray flux begins to increase, while
the optical and UV emission continues to decline.  Here we argue how
the scenario for the X-ray emission of \J16 described in \S\ref{xray}
can reconcile these observations with the X-ray reprocessing origin of
the optical emission.

As shown by Dubus et al. (1999b) the outer regions of a planar
accretion disc cannot intercept the X-rays emitted by a point source
located at the midplane.  Therefore, in order for the outer disc to be
irradiated, it must be warped (the other possible way for the outer
disc to see the X-rays - an extended irradiating source - fails to
explain why the outer disc is {\sl effectively} geometrically thick
while the vertical equilibrium implies very thin discs). 

The origin and propagation of warps in accretion discs is still
an open question (see e.g. Pringle 1999).  Here we consider two
possible regimes: one with low viscosity, when the warp propagation
relies on sound waves; and another with high viscosity, when the warp
evolution is driven by diffusion.  As we shall see, in both cases 
the warp amplitude in \J16 is likely to decay during an outburst.

If the viscosity is low, i.e. if $H/R > \alpha$, where $H$ is the
half--thickness of the disc, the warp can propagate as a
non--dispersive wave at approximately the speed of sound (Papaloizou
\& Lin 1995).  This regime can be relevant if the outer disc regions
are not affected by the propagation of the heat front during the
outburst.  In the standard DIM the heat--front passage changes the
viscosity parameter $\alpha$ from a low, cold ($\sim 0.01$) to a high,
hot ($\sim 0.1$) value.  Therefore, if the heat front does not reach
the outer disc it is conceivable (since we don't know the physical
mechanism supposedly responsible for the change of $\alpha$) that in
the outer regions we would still have $\alpha \sim 0.01$, while $H/R
\gta 0.01$ (see e.g. Dubus et al. 1999b).  In such a case the increased
stream of mass transfered from the secondary deposits matter moving in
the orbital plane at $\sim R_{\rm circ}/2$.  Since various parts of
the disc communicate efficiently through sound waves, this new
component will exert a torque on the outer disc reducing the warp on a
time scale of a few forced precession periods, where the precession
period is $\sim 40$ days (Larwood 1998).

On the other hand, the whole disc could be in a high $\alpha$ state
during an outburst.  Since in a standard accretion disc $H/R \lta
0.05$ in the outer regions, $H/R \ll \alpha$ and the warp propagation
is driven by viscous processes.  The relevant viscosity is the one
corresponding to the vertical shear.  The ratio of this kinematic
viscosity coefficient to the standard (radial) one is approximately
$1/2\alpha^2$, for $\alpha \ll 1$ (Papaloizou \& Pringle 1983), and
therefore the warp damping time is $t_{\rm damp} \approx 4 \alpha^2
t_{\rm vis}$.  At the outer disc edge of \J16, $t_{\rm damp}$ is then
about $100$ days for $\alpha=0.1$, in very good agreement with our
scenario.  Note that in this regime the increase of mass transfer would
not affect the warp.

Pringle (1996) found that warp can be radiation driven. In such a case the
warp's viscous decay could be prevented by irradiation. For this to
happen, the growth rate of the radiative instability must be shorter
than the viscous damping time $t_{\rm damp}$. This condition can be
written as (e.g. Wijers \& Pringle 1999):
\begin{eqnarray}
\gamma_{\rm crit} > 3.21 \left({\eta \over 0.1}\right)^{-1} 
\left({\alpha\over 0.1}\right)^{-2}&&
\left({M \over 7\Msun}\right)^{1/2} \times \nonumber \\
&& \left({R \over 10^{11} {\rm cm}}\right)^{-1/2},
\label{radinst}
\end{eqnarray}
where $\eta$ is the accretion efficiency and $\gamma_{\rm crit}\approx 0.1$ 
is the critical ratio of the radiative growth to the viscous damping times. 
One can
see that for \J16 the inequality above can be satisfied only for
$\alpha$ values higher than the ones usually assumed in the DIM. Unless
such values are assumed a warp will be viscously damped on a time--scale
estimated above.

Whatever the mechanism of warp decay, it would result in the
reduction of the irradiating flux intercepted by the disc, and a
consequent decrease in the observed optical flux from the system.

To illustrate this argument we calculated a series of optical spectra
from uniformly accreting thin discs with varying degree of
irradiation.  The value of the mass accretion rate was chosen to
reproduce the black body component of X-ray emission (Sobczak et
al. 1999).  In our simple treatment here we specify neither the origin
nor the structure
of the warp.  However, since we need some description of the photosphere of
the disc above the orbital plane as a function of radius, we use the
prescription,
\begin{equation}
z = z_{\rm out} (R/R_{\rm out})^{9/7},
\label{z}
\end{equation}
chosen by analogy with formulae used in the literature, which
(despite being based on an incorrect assumption about irradiated
discs) seem to give a correct empirical description of the reprocessed
X--ray flux (see Dubus et al. 1999b).
All other properties of the warped disc, as far as irradiation is
concerned, are described by ${\cal C}$, defined in Eq. (\ref{tirr2}).
We use the prescription for irradiation of the outer disc by the inner
disc edge (e.g. see Shakura \& Sunyaev 1973; King \& Ritter 1998),
which combined with Eq. (\ref{z}) above gives
\begin{equation}
{\cal C} = {\cal C}_{\rm out} 
\left(\frac{z/R}{z_{\rm out}/R_{\rm out}}\right)^2 = 
{\cal C}_{\rm out} \left(\frac{R}{R_{\rm out}}\right)^{4/7},
\label{C}
\end{equation}
where ${\cal C}_{\rm out} = {\cal C} (R_{\rm out})$.  Note that
for the disc--disc irradiation geometry, the strength of
irradiation is quadratic in $z/R$, since it depends on the projections
of both the emitting and irradiated annuli.

Of course the exact dependence of $z$ on the disc radius is important
in determining the shape of the optical spectrum. However, there are
many other highly uncertain quantities in the calculation (e.g. radial
profile of the albedo in the outer disc, angular distribution of the
irradiating flux, details of radiative transfer in the atmospheres of
irradiated discs) which all contribute significantly to the appearance
of the disc in the optical band. In addition, we are using a flared
planar disc `approximation' to describe a warped disc. Since all these
uncertainties are hidden inside our parameter ${\cal C}_{\rm out}$,
the exact choice of $z(R)$ is not very important. 
As far as we are concerned, Eqs. (\ref{z})
and (\ref{C}) simply describe the distribution of the irradiating flux
with radius and {\sl do not} result from some assumed vertical disc
structure. Note especially that we {\sl do not} assume that the
irradiated disc is isothermal or adopt a particular value for the disc
aspect ratio at the outer edge.

The resulting spectra computed for different values of ${\cal C}_{\rm
out}$ are shown in figure \ref{spectra}.  For comparison we also
plotted the dereddened spectra of \J16 observed (in order of
decreasing flux) on May 14, June 8, June 20, June 30, and July 22,
1996 with {\it HST}/FOS red prism.  (Note that these spectra
correspond to the optical and UV data points shown on figure
\ref{ltcrv}.)  Since the contribution from the secondary is quite
significant, especially at later times, we have subtracted from the
data an estimate of the quiescent optical spectrum of \J16, adjusted
according to orbital phase (see Hynes et al.\ 1998 and Hynes 1999 for
a fuller description of data processing).  The broad hump centered at
$\log{\nu} \sim 14.8$ which remains after subtraction in the last two
spectra has a similar spectrum to the secondary, and we suspect is
probably due to a residual contribution from the companion.  This
suggests brightening of the secondary in outburst due to X-ray
heating, supporting our irradiation scenario.

Though the model spectra are not intended as a formal fit to the data
because of many simplifications in our calculation, the agreement
between the five sets of spectra is fairly good.  Figure
\ref{spectra} shows qualitatively that by decreasing the degree of
warping (as described by ${\cal C}_{\rm out}$), and therefore
irradiation, of the outer disc, we can mimic the observed evolution of
\J16 in the optical band.  Note how with decreasing ${\cal C}_{\rm
out}$ the model optical spectra become softer, just as observed, even
allowing for some residual contribution from the secondary in the data.

One should keep in mind that all model spectra shown in figure
\ref{spectra} were computed for a planar disc with $z \propto R^{9/7}$
and a fixed outer radius, given by Eq. (\ref{rdisc}).  This means
varying $\cal{C}_{\rm out}$ in our model corresponds simply to
different values of the photospheric height at the outer radius.  A
more realistic description of the effects caused by enhanced mass
transfer should, of course, include both changes in the disc profile
(different functional form for $z(R)$) as well as possible changes in
the disc outer radius.  We feel that as long as we use a flared planar
disc to approximate the effects of the warp, the exact disc shape is
beyond the scope of this paper.  Similarly, it is difficult to
calculate the change in the outer radius of the disc from first
principles.  However, observations and numerical simulations of
dwarf nova discs (e.g. Smak 1984; 1999b) show that at the onset of the
outburst $R_{\rm out}$ expands by about $30\%$ (from the canonical
$70\%$ to nearly $90\%$ of the Roche radius) for a constant mass
transfer rate from the companion.  By comparison, simulations with
enhanced mass transfer (say by a factor of 30) simply decrease this
value to $\sim 25\%$, so the difference is too small to be constrained
by the data.

\section{End of the outburst}
\label{end}

Because of many uncertainties in the physics of irradiation, it is
very difficult to infer the time scale of the warp decay from the
optical data. In both regimes considered in the previous section,
our estimates of the decay time (a few forced precession times in the
low viscosity case [Papaloizou \& Terquem 1995; Larwood 1998] or the
viscous damping time in a high viscosity case) are of the same order
as the duration of the plateau phase of the outburst. It is
therefore very tempting to speculate that the decay of the warp is not
only responsible for the decline in the observed optical emission, but
can also, by reducing heating through irradiation, terminate the
outburst.  Even if the mass transfer rate from the secondary remains
high, a decrease in the irradiating flux can cause an increase in the
minimum rate required for stable accretion (see Eq. \ref{mcrirr})
which is sufficient to bring the disc back into the unstable regime,
and therefore, partially shut off the accretion in the inner disc
where X-rays are produced.  However, since the disc will most likely
remain close to the stability limit, a very short period of mass
accumulation at the outer edge is required to trigger the next
outburst, just as was observed for \J16 in 1997 (e.g. Sobczak et
al. 1999).

\section{Discussion and Conclusions}
\label{disc}

We have shown that X-ray and optical observations of \J16 during its
1996 outburst can be reconciled with the standard disc instability
scenario under the assumptions that the outer accretion disc is warped
and that the primary `standard' outburst triggers a burst of mass
transfer from the secondary.  In our picture, the enhancement in the
mass transfer rate was sufficient to create a period of stable
accretion, which was observed as a $\sim 5$ month plateau in the
X-ray light curve.  We have argued that this is entirely plausible,
since available evidence is not inconsistent with the mass
transfer rate in \J16 being close to the stability limit of an
irradiated disc.  Thus a rather small enhancement in $\dot M_{\rm T}$
is required to stabilize it.  In addition, a burst of mass from the
secondary is likely to decrease the warp amplitude in the outer disc.
This effect explains why the optical light curve showed a decline as
the X-ray flux was increasing, and can also account for the spectral
evolution of the optical flux.

Warped disc physics is rather complicated and calculating the
thermal--viscous instability in a warped disc (not to mention the
effects of the enhanced mass transfer) is out of the question, at
present.  Because of this, the arguments presented in this paper are
based on the calculations for planar discs and are necessarily of a
qualitative nature.  A lot more work is necessary before we
can venture to make detailed quantitative comparisons between model
predictions and what is observed.

It is also important to emphasize that although the overall ratio of
X-ray to optical flux (determined by the value of ${\cal C}$ at the
outer disc edge) is well constrained by observations, the detailed
optical spectra of irradiated discs depend on many unknown parameters.
While the angular distribution of irradiating flux may in some cases be
deduced from detailed modeling of X-ray spectra, the exact shape of the
disc as well as the X-ray albedo can only be determined from future
theoretical calculations.

Despite these uncertainties, we believe that the picture presented
here is realistic and is consistent with all the available data on the
1996 outburst of \J16.  One should perhaps view our result as a test
of the hypothesis that this outburst was due to a dwarf--nova type
disc instability.  It is important to note that in our arguments we
rely on only two major assumptions, which are both supported by
extensive circumstantial evidence. Disc warping emerges as a natural
conclusion from studies of X-ray irradiated outer discs, and a mass
transfer increase during the outburst seems equally plausible when we
consider irradiation of the secondary. 

Similar modifications are also required in the application of the DIM
to standard (U Gem -- type) dwarf novae.  In these systems there is
clear observational evidence for irradiation--induced enhancement in
the rate of mass transfer from the secondary (e.g. Smak 1995). Smak
(1999a) studied the two types of `inside--out' outbursts which are
observed in dwarf novae: the `narrow' and `wide' ones. He showed, that
their properties (various widths but similar amplitudes) can be
explained if during the outburst the mass transfer rate is increased.
Narrow outbursts correspond to moderate (factor 2) enhancements
whereas wide outbursts correspond to major enhancements which
temporarily put the accretion disc into the stable regime. It is,
therefore, reasonable to expect that the variety of light curve
types observed in X--ray transient systems may be due, at least in part,
to mass transfer variations.

\J16 is an exceptional system but is not the only one showing a `flat
top' outburst light curve (see e.g. Chen et al. 1997). One may hope,
therefore, that such systems could be observed in a not too distant
future. Multi-wavelength observations of such a transient event, in
particular the correlation between X-ray and optical fluxes, could
provide a useful test of our scenario.

At least one feature of the observed behaviour of \J16 still remains a
puzzle.  If the mass transfer rate in quiescence is always close to
the stability limit, it is unclear how this system could have
sustained a $>30$ years quiescence period.  At such a high rate of
accretion, the mass accumulation period at the outer edge of the disc
would be very short (on the order of a few months to a year), with the
successive outbursts following each other after a correspondingly
short time.  In fact, the behaviour observed in this system since 1994
should be the norm, rather than the exception.  One possible
explanation can be that during some periods accretion in \J16 is
non--conservative, e.g. King \& Kolb (1999) propose that a lot of
matter can be lost into a jet.  However, more theoretical work as well
as observations are required before we can assemble a comprehensive
picture of this fascinating system.

\begin{acknowledgements}
We thank Caroline Terquem for enlightening discussions and advice
and Phil Charles for carefully reading the manuscript.
We thank the third referee of this article (first for A\&A) 
for a useful, critical and professional report.
We are grateful to Greg Sobczak for the X-ray light-curve data. This
research was supported in part by the National Science Foundation
Grant No. PHY94-07194, and by NASA through Chandra Postdoctoral
Fellowship grant \#PF8-10002 awarded by the Chandra X-Ray Center,
which is operated by the Smithsonian Astrophysical Observatory for
NASA under contract NAS8-39073.  {\it HST} data was supported by NASA
through grant number GO-6017-01-94A from the Space Telescope Science
Institute, which is operated by the Association of Universities for
Research in Astronomy, Incorporated, under NASA contract NAS5-26555.
We thank Carole Haswell for providing these data.
\end{acknowledgements}

\label{lastpage}


\begin{thebibliography}{}

\bibitem{aks93} Augusteijn, T., Kuulkers, E., Shaham, J. 1993,
\aaa279, L13

\bibitem{csl97} Chen, W., Shrader, C. R., Livio, M. 1997, \apj491, 312

\bibitem{dlh99} Dubus, G., Lasota, J.-P., Hameury, J.-M. 1999a, in preparation

\bibitem{dlhc99} Dubus, G., Lasota, J.-P., Hameury, J.-M., Charles,
P. A. 1999b, \mnras303, 139

\bibitem{fkl87} Frank, J., King, A.R.,  Lasota, J.-P. 1987, \aaa178, 137

\bibitem{hlh97} Hameury, J.-M., Lasota, J.-P., Hur\`e, J.-M. 1997,
\mnras287, 937

\bibitem{hkl88} Hameury, J.-M., King, A.R., Lasota, J.-P. 1988,  \aaa192,
                   187
\bibitem{hlw}   Hameury, J.-M., Lasota, J.-P., Warner, B. 2000, \aaa{353},
244

\bibitem{hkl93} Hameury, J.-M., King, A.R., Lasota, J.-P., Raison,
F. 1993, \aaa277, 81

\bibitem{hlm97} Hameury, J.-M., Lasota, J.-P., McClintock, J. E., Narayan,
R. 1997, \apj489, 234

\bibitem{hmd98} Hameury, J.-M., Menou, K., Dubus, G., Lasota, J.-P.,
Hur\`e, J.-M. 1998, \mnras298, 1048

\bibitem{hjr95} Hjellming, R. M., Rupen, M. P. 1995, \nat375, 464

\bibitem{hea98} Hynes, R. I. et al. 1998, \mnras300, 64

\bibitem{hyn99} Hynes, R. I. 1999, Ph.D. Thesis, The Open University

\bibitem{kic98} King, A. R., Cannizzo, J. K. 1998, \apj499, 348

\bibitem{kik99} King, A. R., Kolb, U. 1999, \mnras305, 654

\bibitem{kir98} King, A. R., Ritter, H. 1998, \mnras293, L42

\bibitem{kol98} Kolb, U. 1998, \mnras297, 419

\bibitem{kkr97} Kolb, U., King, A. R., Ritter, H., Frank, J. 1997, \apj458, L33

\bibitem{lar98} Larwood, J. 1998, \mnras299, L32

\bibitem{mnl99} Menou, K., Narayan, R., Lasota, J.-P. 1999a, \apj513, 811

\bibitem{mel99} Menou, K., Hameury, J.-M.,  Lasota, J.-P., Narayan, R. 1999b
MNRAS, in press

\bibitem{orb97} Orosz, J. A., Remillard, R. A., Bailyn C. D.,
McClintock, J. E. 1997, \apj478, L83

\bibitem{p96} Pringle, J.E. 1996, MNRAS, 281, 357

\bibitem{p99} Pringle, J.E. 1999, in Astrophysical Discs, ASPCS 160, eds. J.A.
Sellwood \& J. Goodman, p. 53

\bibitem{pl95}  Papaloizou, J. C. B., Lin, D. N. C. 1995, \apj438, 841

\bibitem{pap83} Papaloizou, J. C. B., Pringle, J.E. 1983, \mnras202, 1181

\bibitem{pat95} Papaloizou, J. C. B., Terquem, T. 1995, \mnras274, 987

\bibitem{psp99} Phillips, S. N., Shahbaz, T., Podsiadlowski,
Ph. 1999, \mnras304, 839

\bibitem{rtw98} Reg\"os, E., Tout, C. A., Wickramasinghe, D. 1998, \apj509, 362

\bibitem{shc99} Shahbaz, T., van der Hooft, F., Casares, J., Charles,
P. A., van Paradijs, J. 1999, \mnras306, 89

\bibitem{ss73} Shakura, N.I., Sunyaev, R.A. 1973, \aaa22, 41

\bibitem{sma84} Smak, J.S. 1984 \acta34, 161

\bibitem{sma95} Smak, J.S. 1995 \acta45, 355

\bibitem{sma99a} Smak, J.S. 1999a \acta49, 383

\bibitem{sma99b} Smak, J.S. 1999b \acta49, 391

\bibitem{smr99} Sobczak, G. J., McClintock, J. E., Remillard, R. A.,
Bailyn, C. D., Orosz, J. A. 1999, \apj520, 776

\bibitem{tas96} Tanaka, Y. \& Shibazaki, N. 1996, \annrev34, 607

\bibitem{par96} van Paradijs, J. 1996, \apj464, L139

\bibitem{pam95} van Paradijs, J., McClintock, J.E. 1995, in X-ray Binaries,
eds. W.H.G. Lewin, J. van Paradijs \& E.P.J. van den Heuvel, (Cambridge: CUP),
p. 58

\bibitem{wp99} Wijers, A.M.J., Pringle, J.E. 1999, MNRAS, 308, 207

\end{thebibliography}
\end{document}